# Coexistence of Tunable Weyl Points and Topological Nodal Lines in Ternary Transition-Metal Telluride TaIrTe$_4$

Xiaoqing Zhou[1,†,*], Qihang Liu[2,†,*], QuanSheng Wu[3,4], Tom Nummy[1], Haoxiang Li[1], Justin Griffith[1], Stephen Parham[1], Justin Waugh[1], Eve Emmanouilidou[5], Bing Shen[5], Oleg V. Yazyev[3,4], Ni Ni[5], and Daniel Dessau[1]

[1]*Department of Physics, University of Colorado, Boulder, CO 80309, USA*
[2]*Renewable and Sustainable Energy Institute, University of Colorado, Boulder, CO 80309, USA*
[3]*Institute of Physics, École Polytechnique Fédérale de Lausanne (EPFL), CH-1015 Lausanne, Switzerland*
[4]*National Centre for Computational Design and Discovery of Novel Materials MARVEL, Ecole Polytechnique Fédérale de Lausanne (EPFL), CH-1015 Lausanne, Switzerland*
[5]*Department of Physics and Astronomy and California NanoSystems Institute, University of California, Los Angeles, CA 90095, USA*
† These authors contributed to the work equally.
*E-mail: Xiaoqing.zhou@colorado.edu, qihang.liu85@gmail.com

**Abstract**

We report a combined theoretical and experimental study on TaIrTe$_4$, a potential candidate of the minimal model of type-II Weyl semimetals. Unexpectedly, an intriguing node structure with twelve Weyl points and a pair of nodal lines protected by mirror symmetry was found by first-principle calculations, with its complex signatures such as the topologically non-trivial band crossings and topologically trivial Fermi arcs cross-validated by angle-resolved photoemission spectroscopy. Through external strain, the number of Weyl points can be reduced to the theoretical minimum of four, and the appearance of the nodal lines can be switched between different mirror planes in momentum space. The coexistence of tunable Weyl points and nodal lines establishes ternary transition-metal tellurides as a unique test ground for topological state characterization and engineering.

In recent years, topological semimetals (SMs) has emerged as an extremely active topic due to their exotic physics both in bulk and surface states [1-4]. Such SMs host crossing points between valence and conduction bands that are represented by Weyl fermions [2,5-18], Dirac fermions [3,19-22], nodal-lines [1,9,23-28] or even quasi-particles without analogous states in the standard model of high-energy physics [17,29-33]. With the explosion of discovery of these topological SMs, there arises a natural question of which material should be regarded as a prototype of a given class, with the ease to grow and with simple topological features that can be measured. For example, $Bi_2Se_3$ [34] and $Na_3Bi$ [22] have experimentally shown such merits as the benchmarks of topological insulators and Dirac SMs, respectively, while triclinic $CaAs_3$ has been theoretically identified as a minimal "hydrogen atom" model of nodal-line SM recently [35]. Similarly, a minimal approach [18] for the Weyl SM is to find the one with the band structure as clean as possible near the Fermi level ($E_F$) and more importantly, with the smallest number of Weyl points. Given the difficulty to realize an ideal magnetic Weyl SM, the target of the "hydrogen atom" model for the Weyl SMs has been focused on the inversion-breaking SMs with the minimal number of Weyl points equal to four.

Recently, the family of transition-metal chalcogenides $MTe_2$ ($M$ = Mo, W) [6,10-12,14,15] and $MM'Te_4$ ($M$ = Ta, Nb; $M'$ = Ir, Rh) [9,13,16,17,36] have been predicted and experimentally identified to be type-II Weyl SMs, with tilted Weyl cones appearing at boundaries between electron and hole pockets by breaking Lorentz invariance. In $WTe_2$ and $MoTe_2$, the underlying Weyl points are subject to very subtle changes (such as lattice constant) [14,15], posing great difficulties to identify the number of Weyl points. Furthermore, even the benchmark signature, i.e., the Fermi arcs observed in angle-resolved photoemission spectroscopy (ARPES) could often be "fake evidences" of the existence of Weyl points because of the existence of bulk carrier pockets at the Fermi surface [11,14]. Indeed, the best way to confirm the existence of Weyl points and other topological non-trivial features such as nodal lines is to establish a close correspondence between experiment and theory and let the theory judge if the phenomena are of topological origin. On the other hand, $TaIrTe_4$ has been proposed as the minimal model of Weyl SM with only four Weyl points well-separated in $k$-space [13], which was supported by a few experimental reports [9,16]. However, whereas the four Weyl points located ~ 100 meV above $E_F$ could be somewhat supported, there is no evidence that the system does not host any gap closing points other than these four Weyl points. In fact, in both works [9,16] the spectra near $E_F$ and $\Gamma$ point are relatively blurry, imposing a great challenge to clearly identify the subtle band crossings and novel surface states associated with the possible topological features in that area.

In this work, we use the combination of ARPES and first-principles calculations based on density functional theory (DFT) [37-39] to resolve the electronic structure of TaIrTe$_4$, especially focusing on the details very close to $E_F$. In contrast to the previous experimental studies [9,16] that utilized relatively low resolution, we find theoretically that TaIrTe$_4$ hosts a unique combination of twelve Weyl points and two Dirac nodal rings in the Brillouin zone (BZ). Our very high resolution ARPES measurement by 7-eV laser successfully resolved the details close to the Γ point, and found band crossings below $E_F$ as evidence of the bulk nodal lines by comparing with DFT results. We also found apparent surface Fermi arcs that are topologically trivial, indicating the importance of the crosscheck between both theory and experiment to confirm the topological features in type-II Weyl SMs with complex Fermi surfaces. Our observation thus rules out TaIrTe$_4$ as a minimal model of Weyl SM, but establishes it as a promising candidate for topological band engineering instead. Notably, applying moderate strain to TaIrTe$_4$ would tune the number of the Weyl points as well as the appearance of the nodal line within different mirror planes in momentum space.

***Coexistence of Weyl points and nodal lines in TaIrTe$_4$.*** The structure of TaIrTe$_4$ could be viewed as a cell-doubling derivative of WTe$_2$ (along *y* direction), with 1T'-type layer structure and Bernal AB stacking. As shown in the inset of Fig. 1a, within each monolayer, there is Peierls-like dimerization along *y* direction lowering the symmetry, forming short-long-short-long bonds sequence between metal cations and Te anions. With single crystal X-ray diffraction (Fig. 1a) we determined the structure as an orthorhombic lattice with the lattice constants *a* = 3.80 Å, *b* = 12.47 Å and *c* = 13.24 Å, which was used in our DFT calculations [40]. As shown in Fig. 1b, transport study shows clear quantum oscillations and non-saturating magnetoresistance, the latter of which is suggested to be relevant with the existence of Weyl points [41,42]. More details of the synthesis, structural data and transport properties are shown in Supplementary Section S1 [43].

Similar to other type-II Weyl SMs WTe$_2$ and MoTe$_2$, TaIrTe$_4$ has a non-centrosymmetric *Pmn2$_1$* (No. 31) space group. This structure contains four symmetry operations: identity, mirror reflection *M$_x$*, screw axis operation {*C$_{2z}$*|(1/2,0,1/2)}, which is a two-fold rotation about *z* axis followed by a fractional lattice translation, and a glide reflection operation $\mathcal{M}$ = {*M$_y$*|(1/2,0,1/2)}. $\mathcal{M}$ transforms (*x*, *y*, *z*) in position space to (*x*+*a*/2, -*y*, *z*+*c*/2), while $\mathcal{M}^2$ transforms (*x*, *y*, *z*) to (*x*+*a*, *y*, *z*+*c*). We note that in the presence of SOC with the anti-unitary operator *T* behaving as $T^2$ = -1, $\mathcal{M}^2$ rotates spin by 2π, leading to a factor of -1 for a spin-1/2 system. Therefore, we have $\mathcal{M}^2 = -e^{i(k_x+k_z)}$, which means that in the $k_y$ = 0 or π plane that is invariant under $\mathcal{M}$, each band has an eigenvalue of $\mathcal{M}$ either $ie^{i(k_x+k_z)/2}$ or $-ie^{i(k_x+k_z)/2}$. In this inversion-breaking system where every band is singly degenerate, the valence and conduction bands may cross each other along a nodal line if they possess opposite $\mathcal{M}$ eigenvalues. Similarly, $M_x$ mirror symmetry can also support nodal-line structure within the $k_x = 0$ or $\pi$ plane, in the presence of SOC.

The original goals of replacing Mo (or W) with Ta and Ir were 1) to realize a material with only four Weyl points, which is the minimum number of Weyl points for an inversion-breaking Weyl SM; and 2) to extend the separation of Weyl points in momentum space [9,13]. However, our DFT calculations show (see Supplementary Section S2 for the methods [43]) that TaIrTe$_4$ has a more complex structure of topological band degeneracies than previously suggested. Besides the four equivalent Weyl points [W1, coordinates ±(0.199, 0.071, 0) Å$^{-1}$] located in the $k_z = 0$ plane, in agreement with the previous works [9,16], we found another eight equivalent Weyl points closer to the Γ point [W2, coordinates ±(0.079, 0.021, 0.053) Å$^{-1}$] but off the $k_z = 0$ plane. Fig. 1c shows the connection between one set of W1 and W2 points in the DFT band structure. All these Weyl points are of type II, representing the breaking of Lorentz invariance. We note that W1 is located 74 meV above the Fermi level, which is difficult to be directly measured by conventional ARPES method. On the other hand, the eight W2 Weyl points are 71 meV below the Fermi level, accessible by ARPES but having smaller $k$-space separation than that of W1. Interestingly, in TaIrTe$_4$ we also found two nodal lines in the $k_y = 0$ plane in the presence of SOC. As shown in Fig. 1d, the nodal lines do not surround any high-symmetry points, and thus form two symmetric parts with respect to the Γ point in the BZ. Furthermore, the nodal lines are crescent shaped, rather than circle-like found in other reported nodal-line candidates [9,25,26,35,44]. This is caused by the large anisotropy between the in-plane $x$ and out-of-plane $z$ directions. The energy of the nodal lines is about 67 meV below the Fermi level. As described above, the SOC-robust nodal lines in the $k_y = 0$ plane are protected by the glide reflection operator $\mathcal{M}$, with their non-trivial topology verified by the Berry phase calculations [43].

*ARPES measurements of TaIrTe$_4$.* Fig. 2b shows the Fermi surface observed with ARPES under a synchrotron light of 20 eV photon energy. The first-order features, such as the two side pockets contributed by hole carriers, are generally consistent with the DFT calculations of the bulk band (with a Fermi energy offset of -50 meV) as shown in Fig. 2a. The four W1 Weyl points, with their chirality marked in Fig. 2b, fall into these two hole pockets. The *E-k* dispersion along the Γ-X direction is shown in Fig. 2c. We found that the overall features of the spectra are similar to those from the previous work of Haubold *et al.* [16]. Nevertheless, the spectra near $E_F$ for both of these "conventional" synchrotron measurements [16] are relatively blurry when compared to the very small energy and *k*-space separations of the various states, imposing a great challenge to clearly identify the subtle band crossing and the surface state associated with possible topological features. To overcome this, we performed measurements in this particular area (indicated by black frame in Fig. 2c) using a 7-eV laser, with the observed spectra revealing far more details than those observed using higher photon energies (as shown in Fig. 2e and Fig. 2f). This difference is due to the markedly improved energy and momentum resolution enabled by low-energy laser ARPES [45], which provides a better means for resolving the subtle band behavior like the predicted nodal lines close to the Γ point. In Fig. 2e, the two side pockets at $k_x = \pm 0.199$ Å$^{-1}$ are not visible in the Fermi map, but the spectra near the center demonstrate several sharp lines representing the Fermi surface. More importantly, the calculated W2 Weyl points as well as the nodal lines are all located in this narrow region (see Fig. 2e), providing a much better chance to find the evidence of these topological features. As shown in Fig. 2f, the dispersion spectra are also no longer the blob of intensity as in Fig. 2c, making it possible to identify topological signatures. Specifically, we found two curved branches near the Γ point from $E_F$ to about 40 meV below. Such unconventional band dispersions are not revealed by the global but rough ARPES image in Fig. 2c. The sharp contrast between the 7-eV spectra and the higher photon energy ones are most likely due to the matrix element effect, since the beam spot sizes (less than 40 by 100 μm) are comparable in both cases. It also seems that photon energy plays a far more important role here than the light polarization, as varying the polarization in both cases did not substantially modify the spectra (see Supplementary Section S3 [43]).

To verify the curved feature in Fig. 2f, we next look more closely into surface dispersion by ARPES and connect it to the calculated counterpart. The momentum-resolved local density of states (LDOS) obtained from DFT calculations for the top [001] surface is shown in Fig. 3a (the bulk momentum-resolved DOS is shown in Supplementary Fig. S3a for comparison [43]). We first focus on the $k_y = 0$ cut, where both W2 Weyl points and nodal lines could in principle be resolved. Compared with bulk state, the surface LDOS from Fig. 3a suggests that as the surface band merges into the bulk band, only the left band crossing point at $k_x = 0.07$ Å$^{-1}$ is clearly presented. Our ARPES observation fits the calculated surface dispersion very well (as shown in Fig. 3b), with only the bands on the left of the first crossing point of the nodal line visible. In contrast, the W2 Weyl points overlaps with the second band-crossing points, and so their signatures are likely invisible under 7-eV light. Overall, such a good agreement between DFT and ARPES confirms that TaIrTe$_4$ is not the "hydrogen atom" model for Weyl semimetals; instead, it hosts a 0+1D combination of the nodal surface that is unique among the numerous topological semimetals discovered to date.

Since the type-II Weyl point can be considered as the boundary of electron and hole pocket, generally this kind of Weyl SM hosts a relatively complicated Fermi surface, contributed by both electron and hole pockets. As a result, some topologically trivial surface states could emerge from the bulk continuum, behaving like Fermi arcs. Such trivial Fermi arcs happen in both $MoTe_2$ and $WTe_2$, as reported by previous studies [10,14], and also in $TaIrTe_4$. For example, Fig. 3c and 3d show the theoretical and experimental dispersion spectra below $E_F$ at the $k_y$ = 0.25 Å$^{-1}$ cut. Comparing with Fig. 3c (also see Supplementary Fig. S3b for bulk state [43]), we find that the parabolic electron pocket in the bulk LDOS (Fig. S3b) is truncated in the surface LDOS (Fig. 3c), indicating that only the left part of the surface state emerges from the bulk forming a trivial Fermi arc. The ARPES spectra shown in Fig. 3d show good agreement with the theoretical prediction. Particularly notable in the spectra is a dispersion that has $k_x$ = - 0.065 Å$^{-1}$ and terminates at -8 meV, which resembles a half of a parabolic band and thus seems like a Fermi arc. However, from the verification by DFT calculations we conclude that this appearance is merely the result of a surface band merging into the relatively invisible bulk band at this photon energy. We also note that in Weyl SMs with relatively complex Fermi surfaces, the experimentally observed Fermi arc cannot serve as the direct evidence of the existence of Weyl point or the underlying topological characters [14].

For nodal line SMs, it is predicted that a "drumhead" surface state will form along the nodal-line band crossing points. However, although the nodal lines protected by specific crystal symmetry indeed lead to a nontrivial Berry phase (e.g., $\pi$ in $TaIrTe_4$), such a Berry phase is not sufficient to form a topologically protected surface state. In other words, the expected "drumhead" surface state could be adiabatically deformed into the bulk continuum without any symmetry breaking. Similar situation appears in Dirac SMs where the Fermi arcs on the surface are not topologically protected and can be continuously deformed into the case of a topological or normal insulator [46]. In addition, since the nodal lines are located within the $k_y$ = 0 plane in $TaIrTe_4$, the drumhead surface state is expected to happen at the [010] surface, which is inaccessible for this layered material stacking along the [001] direction. Therefore, we do not expect to verify the existence of nodal lines in $TaIrTe_4$ through surface states.

***Tuning the topological band degeneracies by strain.*** Modifying the number of Weyl points by external "knobs", such as strain, is of practical importance as it may allow achieving a simple Fermi surface with only the minimum number of Weyl nodes, and thus could offer a clearer signature of various topological properties such as the chiral anomaly [47]. On the other hand, the nodal line robust to SOC requires specific mirror symmetries ($M_x$ and $\mathcal{M}$) as mentioned above, and thus appears only within the $k_x$, $k_y = 0$ or $\pi$ mirror plane. Therefore, the location of the nodal lines could also be tuned by changing the lattice parameters, while such interesting manipulation has not yet been reported anywhere. Here we apply compressive and tensile to bulk TaIrTe$_4$ causing its volume $V$ to vary within the range of $\alpha = \frac{V-V_0}{V} = \pm 15\%$, where $V_0$ is the volume under ambient pressure. We first focus on the Weyl points. Within the range of $\alpha$ considered, the four W1 Weyl points within the $k_z = 0$ plane always exists (see Fig. 4a), while other Weyl points appear or annihilate with different lattice constants (see Supplementary Fig. S4 [43]). When $\alpha$ is 12% or 15%, only W1 Weyl points exist, indicating relatively simple node structure. In the case of -12% ≤ $\alpha$ ≤ 6%, there are 12 Weyl points in total. Notably, for compressive strain the 12 Weyl points consist of 4 W1 and 8 equivalent W2 off the $k_z = 0$ plane, same as the case of ambient pressure, while for $\alpha = 6\%$ there are three types of non-equivalent Weyl points all located within the $k_z = 0$ plane.

Whereas the topology of nodal lines remains unchanged under compressive strain, it evolves dramatically with the volume expansion, as shown in Fig. 4b. The two crescent nodal lines within the $k_y = 0$ plane disappear at $\alpha = 6\%$, while two nodal lines within the $k_x = 0$ plane, protected by $M_x$, are created. Such a nodal-line transition under a reasonable volume change is unique among the currently reported nodal-line SMs because of the complex Fermi surface and the two perpendicular mirror planes of TaIrTe$_4$. The nodal lines within the $k_x = 0$ plane still exist for $\alpha = 12\%$ but disappear at $\alpha = 15\%$. Moreover, a new pair of nodal lines within in the $k_y = \pi$ plane appear in both cases. We note that for the case $\alpha = 12\%$ and 15% the node structure is composed by 4 Weyl points and nodal lines with the energy close to $E_F$, which is desirable for ARPES measurement. The energy separation between the Weyl points and nodal lines is about 15 meV. Therefore, we could expect a relatively simple Fermi surface for examining either Weyl or nodal-line topological features by tuning the doping levels in the same material.

In conclusion, by using the combination of ARPES measurement and DFT calculations we find that TaIrTe$_4$ possesses a complex node structure with twelve Weyl points and two crescent nodal lines in the BZ, as opposed to a minimal model of Weyl SM. The delicate entanglement between conduction and valence bands also offers possibility to tune the topology of the Weyl points and nodal lines by moderate strain. By observing topologically trivial Fermi arcs at surface, this work also highlights the importance of the cross-validation between both theory and experiment to confirm the topological features in such type-II Weyl SMs.

**Acknowledgements**

We acknowledge Dr. Y. D. Chuang, Dr. J. D. Denlinger and Dr. M. Hoesch for technical assistances in ARPES, Hao Zheng and Hengdi Zhao for technical assistance in structural determination. This work was carried out with the support of Diamond Light Source beamline I05 (proposal no. SI13406). This work was supported by the U.S. Department of Energy (DOE), Office of Science, Office of Basic Energy Sciences under Award Numbers DE-FG02-03ER46066 (University of Colorado) and DE-SC0011978 (UCLA). This work used resources of the National Energy Research Scientific Computing Center. The Advanced Light Source is supported by DOE office of Science User Facility under contract No. DE-AC02-05CH11231. QSW and OY acknowledge support by the NCCR Marvel.

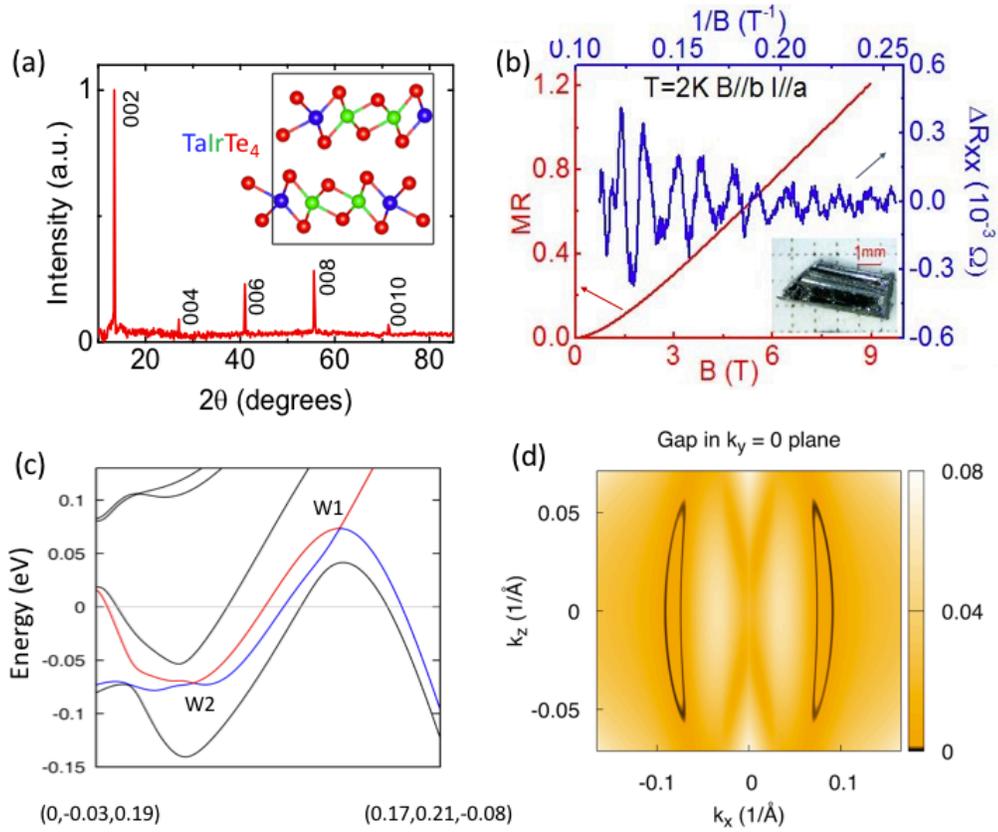

Fig. 1: (a) XRD patterns and crystal structure (inset) of TaIrTe$_4$. (b) Magnetoresistance (red) and Shubnikov-de Haas oscillations (blue). Inset: A single crystal of TaIrTe$_4$ against the 1-mm scale. (c) Band structure plotted along the path from (0, -0.03, 0.19) to (0.17, 0.21, -0.08) in the units of reciprocal lattice vectors. Both W1 and W2 appear along this path. The red and blue lines indicate the conduction and valence band, respectively. The grey line marks the Fermi level. (d) Energy gap (in units of eV) between valence and conductions bands in $k_y = 0$ plane, illustrating two crescent nodal lines.

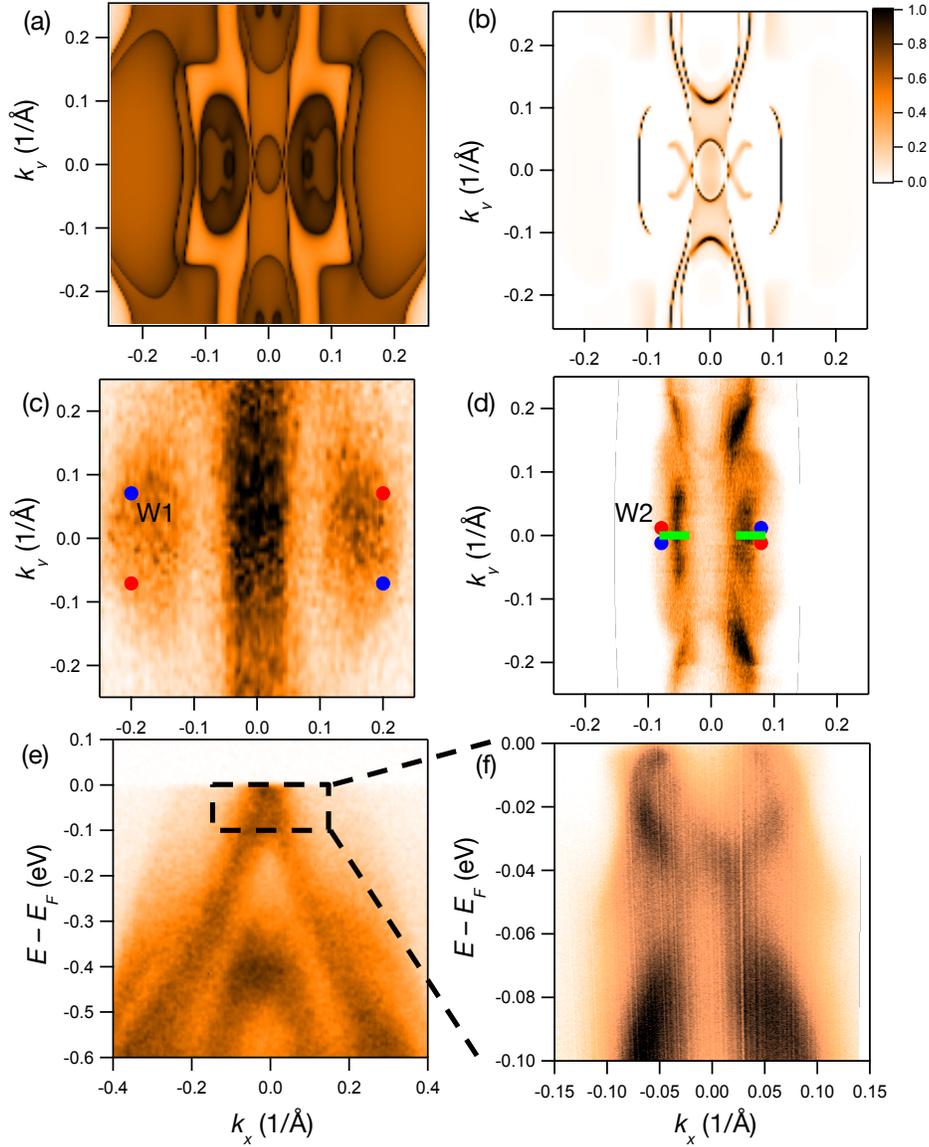

Fig. 2 (a) Momentum-resolved bulk density of states of TaIrTe$_4$ and (b) the corresponding surface-state-only counterpart calculated using DFT at energy 50 meV below $E_F$. (c,d) Experimental ARPES spectra of the Fermi surface taken with (c) 20-eV light, and (d) 7-eV laser. The calculated Weyl points are marked by red (Chern number +1) and blue dots (Chern number -1), whereas the nodal lines projected onto the k$_x$-k$_y$ plane are represented by the green lines. (e,f) ARPES spectra along Γ-X (arrow) for the (e) 28-eV light and (f) 7-eV laser respectively. The black box in panel (e) is zoomed in with more detail in panel (f).

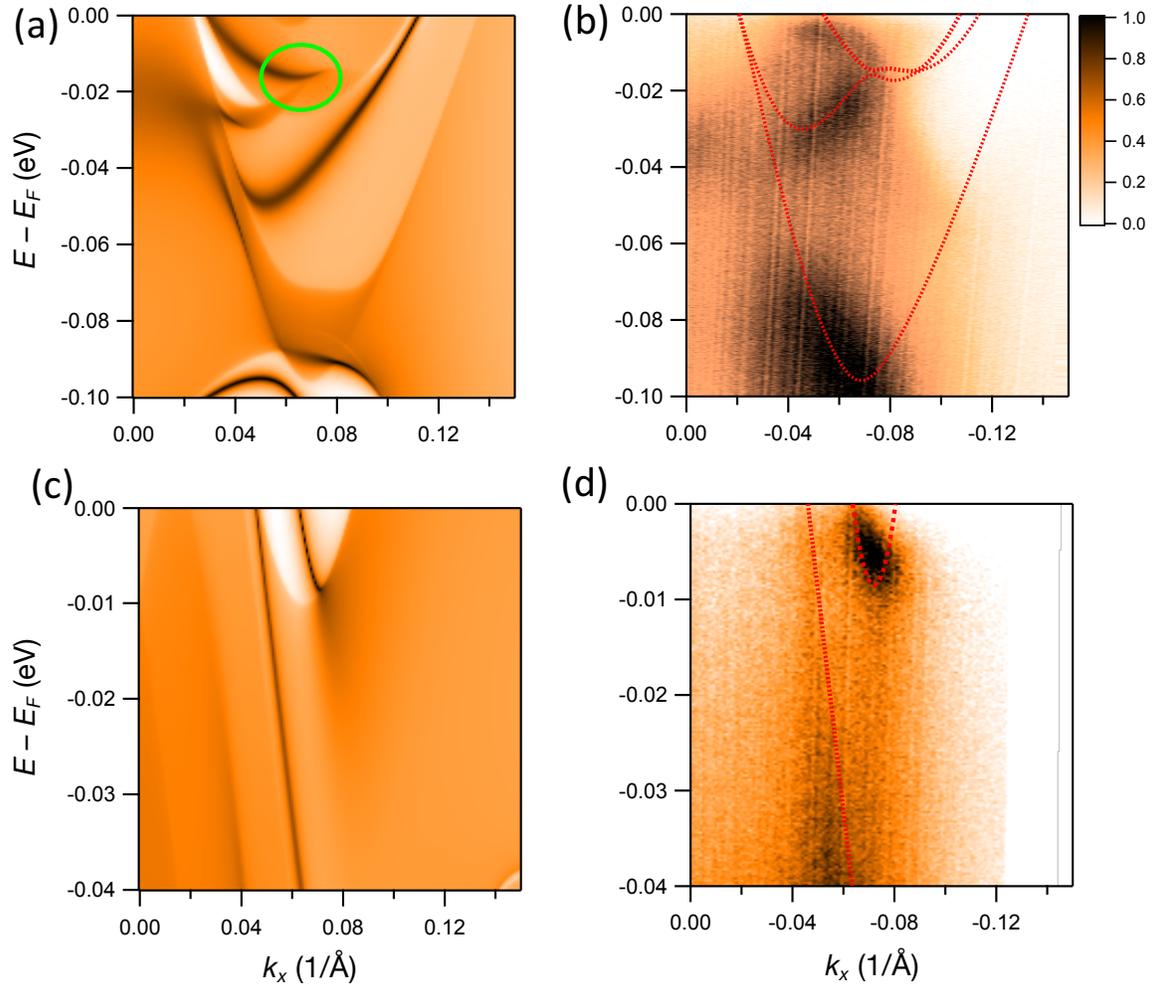

Fig. 3 (a) Momentum-resolved local density of states of TaIrTe$_4$ for the [001] surface at $k_y = 0$ obtained from DFT calculations, to be compared with (b) experimental ARPES spectra taken with 7-eV laser. (c,d) Same as (a,b) but for the cut at $k_y = 0.25$ Å$^{-1}$. The red dashed lines in (b) and (d) are guide-to-the-eye from DFT results for comparison.

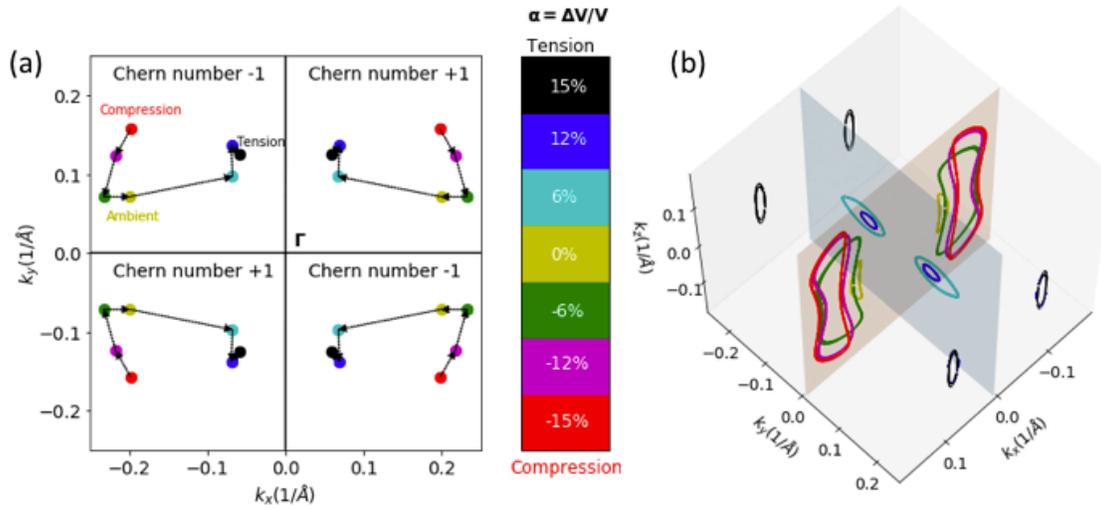

Fig. 4: (a) Evolution of Weyl points W1 upon volume change from $\Delta V/V \sim -15\%$ (red) to $15\%$ (black). (b) Nodal lines switching from the $k_y = 0$ mirror plane (red shade) to $k_x = 0$ (blue shade) and $k_y = \pi$ mirror planes with volume increase.